\documentstyle[sprocl]{article}
\input{epsf}
\def\be{\begin{equation}}
\def\ee{\end{equation}}
\def\bea{\begin{eqnarray}}
\def\eea{\end{eqnarray}}

\begin{document}
\hfill{ANL-HEP-CP-96-64}
\vskip 0.1cm
\hfill{August 20, 1996}

\title{PRODUCTION OF CHARM WITH A PHOTON at $p\bar{p}$ COLLIDERS
\footnote{Talk presented by L. E. Gordon at the DPF96 Meeting of the American 
Physical Society, Minneapolis Minnesota, 10-15 August 1996.}
}
\author{B. Bailey }
\address{Physics Department, Eckerd College, St. Petersburg, FL 33711, U.S.A.}
\author{ E. L. Berger and L. E. Gordon }
\address{High Energy Physics Division, Argonne National Laboratory,\\ 
Argonne, IL 60439, U.S.A.}

\maketitle
\author{B. Bailey} 
\abstracts{
The two particle inclusive cross section for the reaction $p 
+\bar{p}\rightarrow \gamma + c + X$ is studied in perturbative quantum
chromodynamics at order $O(\alpha ^2_s)$. Differential distributions 
are provided for various observables, and a comparison is made with
preliminary data from the CDF collaboration.}
\section{Introduction}

The CDF Collaboration \cite{cdf} are analyzing data on prompt photon 
production in
association with charm decay products such as $e^\pm$, $\mu^\pm$ or
$D^\pm$.  This two-particle inclusive reaction is particularly interesting 
because it offers the possiblility of a detailed study of the underlying QCD 
dynamics such as, e.g., rapidity correlations. In addition the data may 
provide a direct measurement of the charm quark density of the proton due to 
the dominance of the $cg$ scattering process.

We completed two next-to-leading order perturbative QCD calculations 
of the reaction $p +\bar{p}\rightarrow \gamma + c + X$ at high energy 
\cite{bg,bbg}
. In these calculations two different techniques were used in performing the 
phase-space integrals. In the first, purely analytical techniques were used.  
In the second approach, we used a combination of analytical and Monte Carlo 
techniques, which is more flexible and allows implementation of isolation cuts
and other experimentally relevant selections. To warrant use of 
perturbation theory and the massless charm approximation, we limited our 
considerations to values of transverse momenta of the photon and charm quark 
$p^{\gamma,c}_{T} > 10$ GeV. In these calculations we consider and
include hard scattering processes in which the charm quark is a
constituent of the incident hadrons e.g., $cg\rightarrow \gamma c X$.
Details of these calculations can be found
in the above references. 

We recently completed a different calculation of this reaction in which we 
included a finite charm quark mass explicitly. This calculation, which should 
be  valid for $p_T^c\leq 10$ GeV, was done in leading order QCD 
($O(\alpha_s^2)$). In this case the only subprocesses
contributing to the reaction are those in which the
charm quark is produced in the hard scattering reaction, namely
$gg\rightarrow c\bar{c}\gamma$ and $q\bar{q}\rightarrow c\bar{c}\gamma$. 
There are no collinear or soft singularities at this order and no
fragmentation contributions to the reaction. The integration over the
final state phase space could therefore be done numerically with Monte
Carlo routines and photon isolation cuts could be implemented.   

\section{Numerical Results}
Our results are presented at a center-of-mass energy $\sqrt{s}=1.8$ TeV.
Renormalization/factorization scales are taken as $\mu=p_T^\gamma$. 
We sum over charm and anticharm production throughout. Some results are 
presented as distributions in the variable $z=\vec{p_T^c}.\vec{p_T^\gamma}/
(p_T^\gamma)^2$, for finite bin widths of $z$. 
\begin{figure}
{\hskip 0.3cm}\hbox{\epsfxsize5.5cm\epsffile{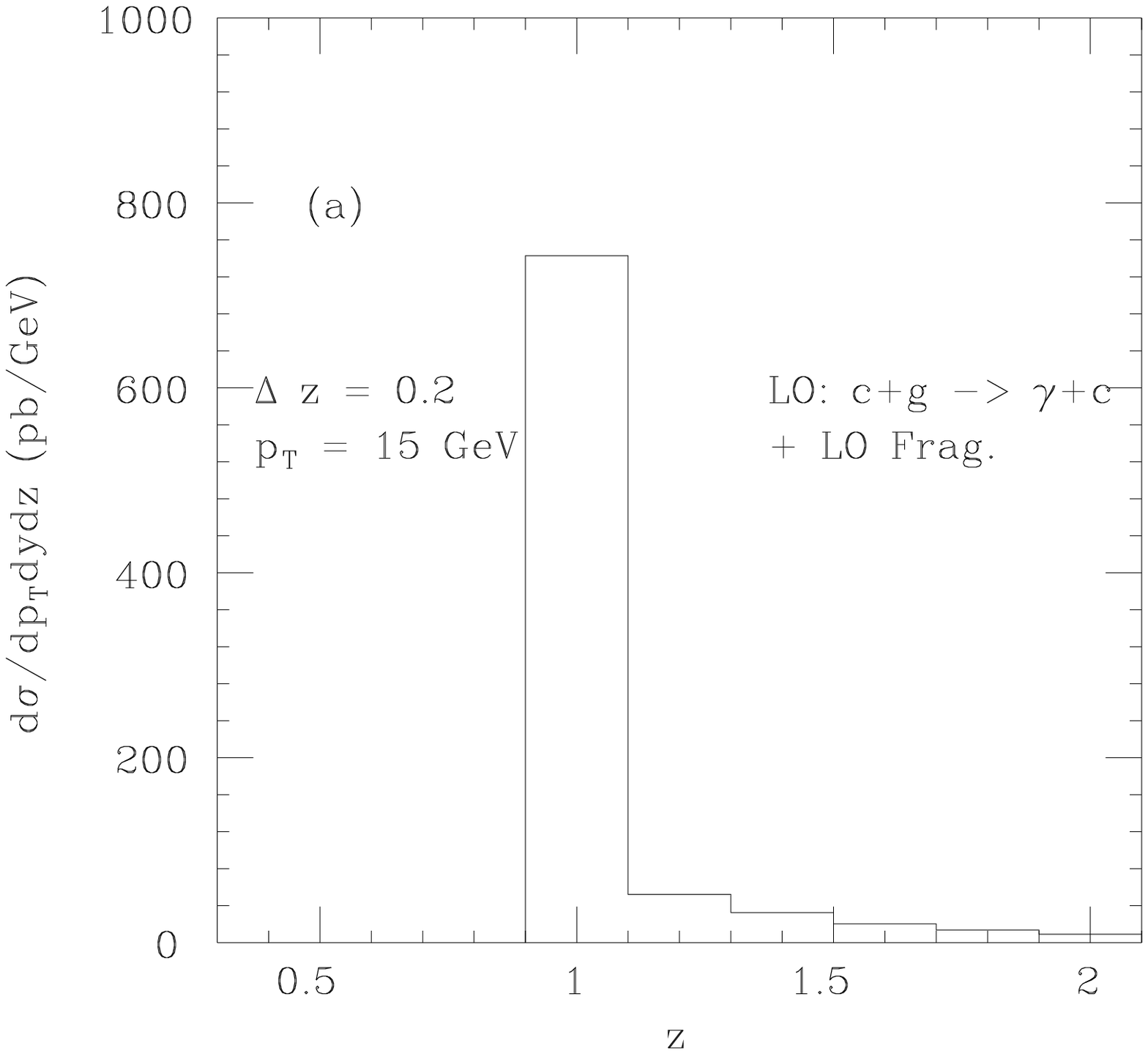}{\hskip 0.2cm}
\epsfxsize5.5cm\epsffile{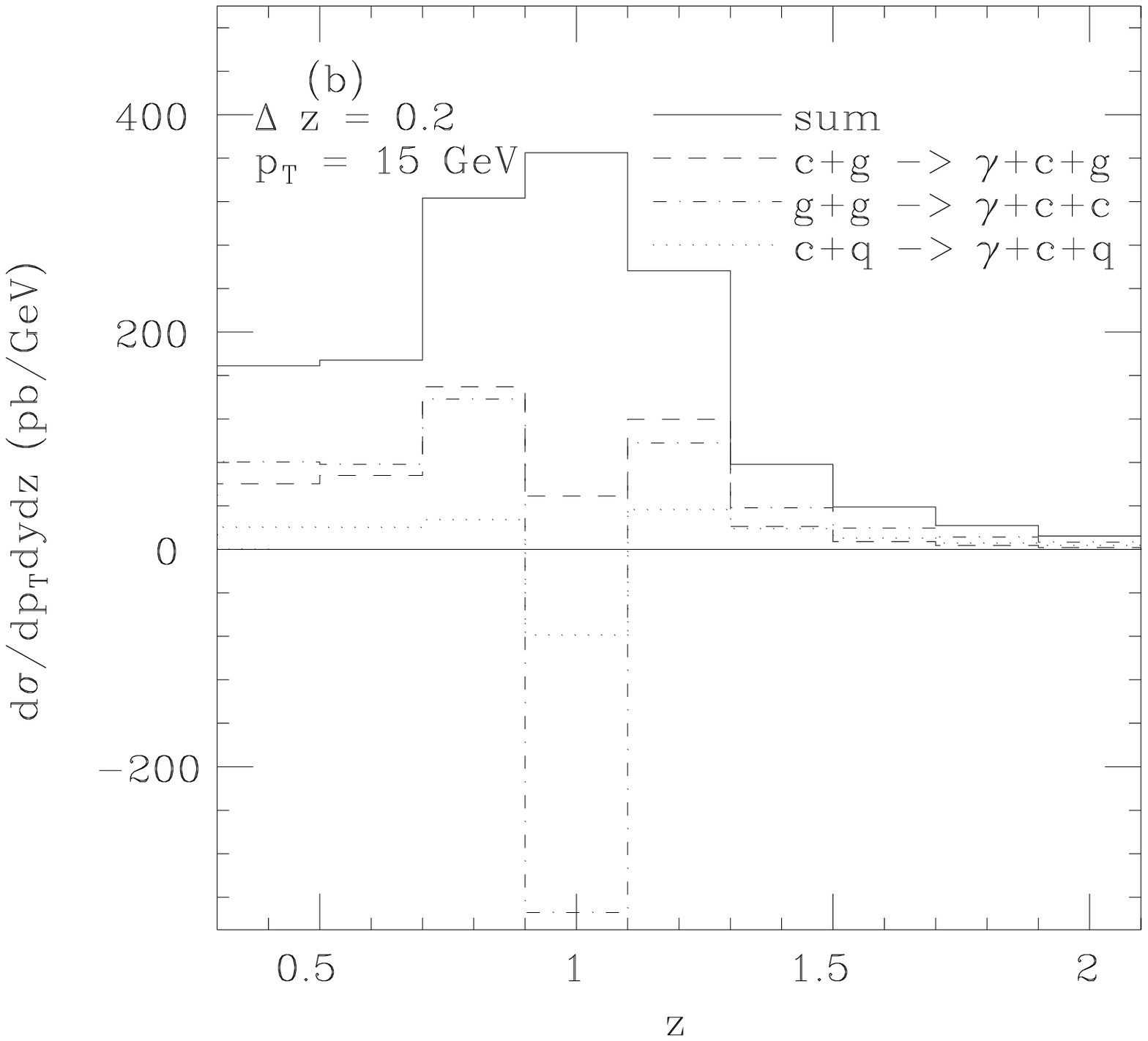}}
\caption{Cross section as a function of z at $y^\gamma=0$ in (a)
leading order and (b) next-to-leading showing the dominant
contributions.}
\end{figure}
In Fig.1a the net lowest order contribution for the massless calculation
is shown as a function of $z$
at $p_T^\gamma=15$ GeV and for a bin size $\Delta z=0.2$. The lowest
order cross section is made up of the lowest order direct term
$cg\rightarrow\gamma c$, which is proportional to $\delta(1-z)$ and
provides the peak at $z=1$, plus the various photon fragmentation
contributions which contribute in the region $z\geq 1$. The most
significant feature of this curve is that there is no contribution to
the cross section in the region $z\leq 1$. This 
unrealistic prediction shows the inadequacy of the lowest order
predictions.

Fig.1b shows the distribution in $z$ predicted by the next-to-leading 
order massless calculation. The next-to-leading order contributions serve to 
lower the 
peak at $z=1$, and they broaden the distribution. The cross section is finite 
at all values of $z$, closer to the
situation observed in experiments. In addition, in Fig.1b we
display contributions from the most important subprocesses. 
The $cg$ initiated process dominates the cross section, but there are important
contributions from the $gg$ and $cq$ initiated process in the low $p_T^\gamma$ 
region. Predicted distributions in $p_T^c$, $p_T^\gamma$ and rapidities
may be found in \cite{bg,bbg}.

In Fig.2a we show the $z$ distribution for the same kinematic
variables as above but now for the leading order massive calculation
with various choices for the charm quark mass. For comparison the full
massless next-to-leading order results are also displayed. Although the massive
calculation is done in leading order, the contributing subprocesses are 
two-to-three scattering processes, and they contribute in the region $z < 1$.
In this sense the massive calculation gives a more realistic result than the 
corresponding massless calculation. On the other hand it clearly predicts a
smaller cross section than the next-to-leading order massless calculation. 
We note, however, that the next-to-leading order contributions to heavy quark
production are known to provide significant increases in the predicted
yields \cite{sally}. Fig.2a 
also demonstrates the sensitivity to the choice for the charm mass 
particularly in the region around $z=1$. This is the 
region in which the finite charm mass regulates the initial state collinear 
divergences.

In Fig.2b we compare our results to CDF data \cite{steve} for
photon plus $\mu^\pm$ production. The three upper points are
obtained\cite{steve} from the Monte Carlo event generator Pythia 
whereas the lower
point is that given by our theoretical calculation. The Pythia cross
sections
lie substantially below the data whereas our cross section exceeds the
data but somewhat closer to it.  
\begin{figure}
{\hskip 0.3cm}\hbox{\epsfxsize5.5cm\epsffile{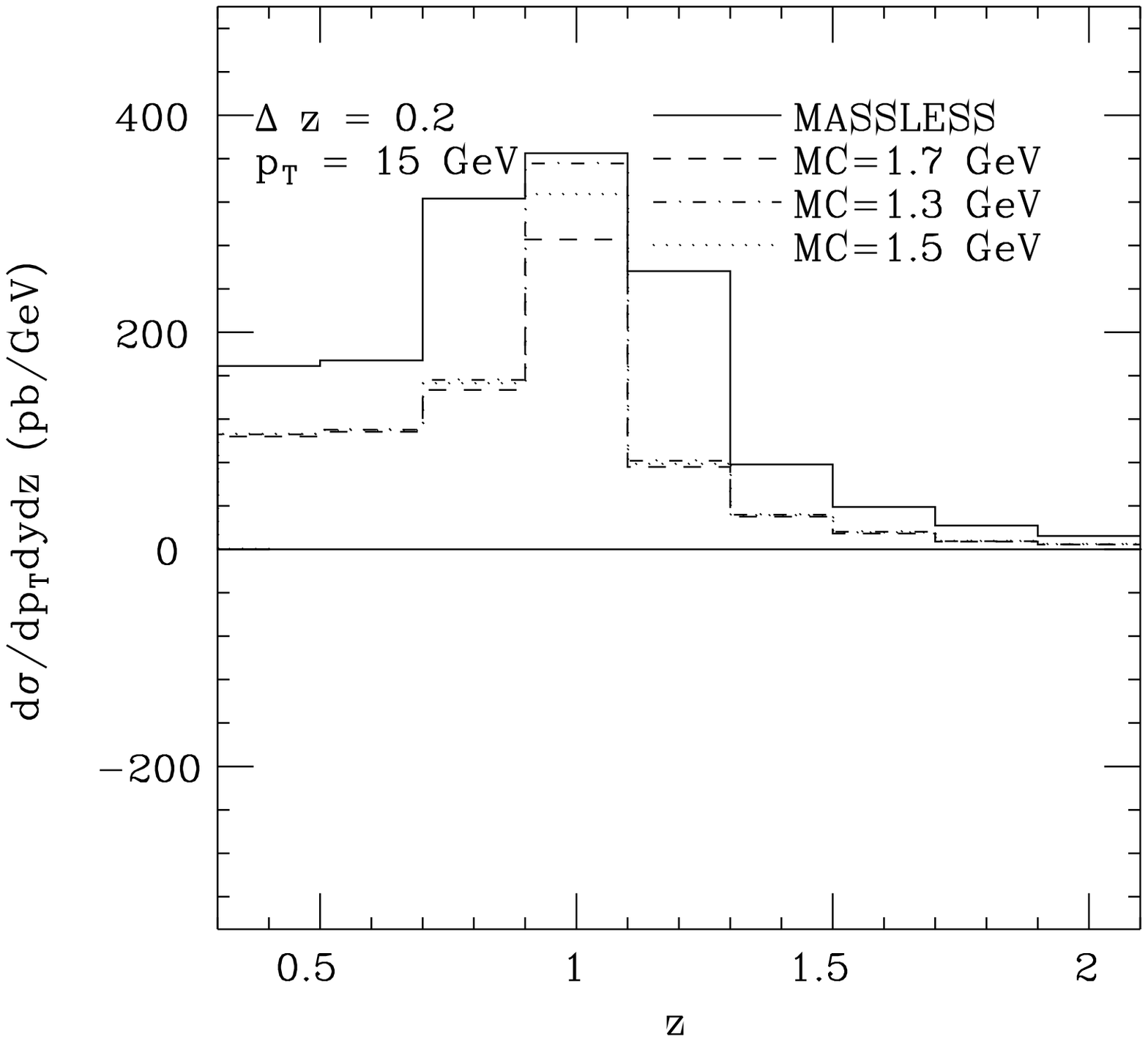}{\hskip 0.2cm}
\epsfxsize5.5cm\epsffile{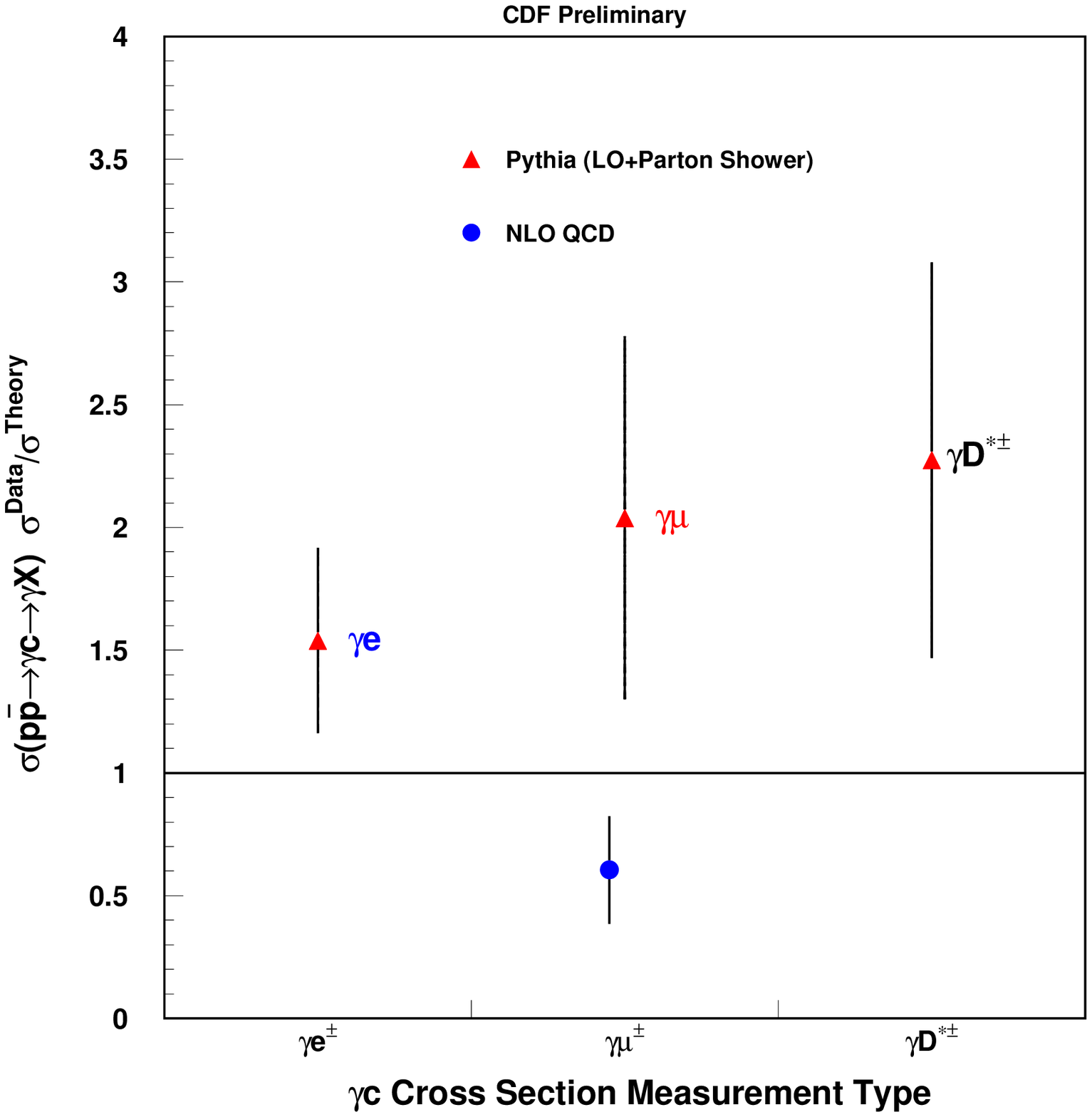}}
\caption{ (a) $z$ distribution of the leading order massive calculation
for various values of the charm mass. The next-to-leading order massless
results are also shown. 
(b) Ratio of the measured cross section to that predicted by theory for
various final state charm decay products.}
\end{figure}

\section{Conclusions}

We presented the results of three calculations of
the inclusive production of a prompt photon in association with a heavy
quark. Two analyses are done at
next-to-leading order in perturbative QCD in the massless charm
framework, and one calculation in leading order QCD in the massive charm
framework.  
Our results in the massless cases agree quantitatively, as they should, but 
the combination of analytic and Monte Carlo methods is more
versatile.  
A comparison of our next-to-leading predictions with the preliminary 
CDF data shows reasonable agreement. 

\section*{Acknowledgments} 
The work at Argonne National Laboratory was supported by the US Department of
Energy, Division of High Energy Physics, Contract number W-31-109-ENG-38.
This work was supported in part by Eckerd College.

\end{document}